# Ejecta Exchange,
# Color Evolution in the Pluto System,
# and
# Implications for KBOs and Asteroids with Satellites

## S. Alan Stern
**22 May 2008**

Visiting Scientist
Lunar and Planetary Institute
3600 Bay Area Blvd.
Houston, TX 77058

alan@boulder.swri.edu

# Abstract

We examine the ability of impacts by Kuiper Belt debris to cause regolith exchange between objects in the Pluto system. We find that ejecta velocities from KB impacts are too low to escape from Pluto and Charon. However, ejecta can escape Nix and Hydra, and is capable of covering one another to depths as high as 10s of meters, and Charon and Pluto, perhaps to depths up to several 10s of cm. Although Pluto's annual atmospheric frost deposition cycle will cover such imported debris on timescales faster than it is emplaced, no such masking mechanism is available on Hydra, Nix, and Charon. As a result, ejecta exchange between these bodies is expected to evolve their colors, albedos, and other photometric properties to be similar. We examined the ability of ejecta exchange to work for other Kuiper Belt binaries and found the process can be effective in many cases. This process may also operate in asteroid binary systems.

# Introduction

Pluto's three satellites, Charon, Nix, and Hydra (Weaver et al. 2006) share neutral colors that are quite unlike the red color of Pluto itself (Stern et al. 2006; see also Table 1 here). As Table 1 illustrates, the colors of Nix and Hydra are indistinguishable.

Table 1: Some Properties of Pluto and Its Satellites [a]

| Object | Absolute Magnitude | B-V Color | Satellite Semi-Major Axis |
|---|---|---|---|
| Pluto | -1.0 | +0.868 | N/A |
| Charon | +0.9 | +0.710±0.011 | 19570 km |
| Nix | +8.64±0.09 | +0.653±0.026 | 49242 km |
| Hydra | +8.48±0.10 | +0.654±0.065 | 65082 km |

[a] Orbits from Tholen et al. (2007); colors from Stern et al. (2007).

This, and their close association with Charon's neutral reflectivity have been cited (e.g., Stern et al. 2006) as primordial evidence that all of these bodies resulted from the giant impact thought to have created the satellite system (McKinnon 1989; Stern, McKinnon, and Lunine 1997; Ward & Canup 2006). In what follows we examine a different hypothesis: that the similar colors of these three bodies are instead the evolutionary result of material exchange between them, owing to the erosive effects of Kuiper Belt bombardment on Nix and Hydra.

## The Effects of Kuiper Belt Bombardment on Pluto's Satellites

Pluto and its satellites, being embedded in the Kuiper Belt, are subject to impacts by Kuiper Belt Objects (KBO) debris. Estimates of the number of impacts and the amount of ejecta that each body in the Pluto system experiences (Durda & Stern 2000; hereafter DS00) give an ejecta generation rate of $2 \times 10^9 (R/100 \text{ km})^2$ gm yr$^{-1}$ for objects in a Pluto-like heliocentric orbit, where R is the radius of the target body. Over the age of the solar system, this corresponds to $8 \times 10^{18} (R/100 \text{ km})^2$ gm of ejecta generation. The DS00 estimate is conservative in that it only considered impactors with radii >4 meters, and therefore ignored the contribution due to still smaller impactors and dust.

KB collisions on objects in the Pluto system characteristically occur at speeds of $V_{imp}$=1-2 km s$^{-1}$. At such speeds, very little vaporization of either $H_2O$-ice or refractory silicates takes place (e.g., Pierazzo et al. 1997). Experimental studies of impacts onto regoliths for impact velocities in this range (e.g., Hartmann 1985) find ejecta typical velocities that are 1-10% of impact velocities, with lower mass ejecta moving at the faster speeds in this range and higher mass ejecta moving at lower speeds in this range. Hence, in the Pluto system, we expect characteristic ejecta speeds of 0.01 km s$^{-1}$ to 0.2 km s$^{-1}$.

Such characteristic ejection velocities are all well below Charon's escape speed, which for a 606 km reference radius and a 1.63 gm cm$^{-3}$ reference density (Tholen et al. 2007), is 0.58 km s$^{-1}$. Therefore, Charon will typically retain ejecta created in impacts. With an even higher escape velocity of 1.3 km s$^{-1}$, Pluto will also typically retain ejecta in collisions. As such, Pluto and Charon are not expected to exchange ejecta with other bodies in the system.

But now consider Nix and Hydra. Although their radii and densities are not known, they can safely be assumed to be <80 km (e.g., Weaver et al. 2006) and <3 gm cm$^{-3}$, respectively, giving a safe upper limit escape speed of 0.1 km s$^{-1}$. More likely radii near 40 km and densities of 1 gm cm$^{-3}$ give an escape speed of 0.03 km s$^{-1}$. Such escape speeds are smaller than the upper range of ejecta velocities that KB impactors should produce. As a result, KB impacts should cause Nix and Hydra to lose regolith material into the Pluto system.

What is the fate of regolith ejecta lost from Nix and Hydra? Heuristically, one can predict that this material will contribute to time-variable ring/dust structures in the system (Stern et al. 2006). A primary loss process for such dust is re-accretion onto one or more of the satellites, or Pluto itself.

Nix and Hydra can efficiently exchange eject with one another or other bodies in the system when the ejecta speed is (i) higher than the minimum speed needed to escape themselves and achieve Pluto-centric orbits with extrema at or beyond the orbit semi-major axis of other bodies in the system, but (ii) lower than the characteristic speed required to escape these bodies and the Pluto system altogether[1].

Table 2 presents calculations of these two speeds for various cases in the Pluto system. We note that the velocities in Table 2 were computed including an assumed, conservative, 45 deg plane change.

Table 2: Some Characteristic ΔVs in the Pluto System

| From-To | Exchange ΔV | Hyperbolic Escape ΔV |
|---|---|---|
| Nix-Pluto | 0.15 km s$^{-1}$ | 0.16 km s$^{-1}$ |
| Hydra-Pluto | 0.14 km s$^{-1}$ | 0.15 km s$^{-1}$ |
| Nix-Charon | 0.11 km s$^{-1}$ | 0.16 km s$^{-1}$ |
| Hydra-Charon | 0.11 km s$^{-1}$ | 0.15 km s$^{-1}$ |
| Nix-Hydra | 0.10 km s$^{-1}$ | 0.16 km s$^{-1}$ |

Note that in all these cases, the exchange ΔV is lower than the hyperbolic escape ΔV, and the hyperbolic escape speed is relatively near the maximum expected ejecta speed. Also note that the velocity window between these two speeds is much narrower in the case of transfers to Pluto than between the satellites in the system.

---

[1] Actually, hyperbolic ejecta can also re-impact other satellites if a satellite happens to fortuitously lie along the escape asymptote, but this process is considerably less efficient than re-impact from trapped orbital debris. The role of hyperbolic ejecta for other KBO and asteroid binaries will be discussed in the next section.

Combining the results shown in Table 2 with the expected 0.01-0.2 km s$^{-1}$ characteristic ejecta speeds, we conclude that Nix and Hydra can exchange material ejected in KB impacts between one another, as well as to Charon and even Pluto.

Because orbital speeds in the Pluto satellite system are just 0.1-0.2 km s$^{-1}$, characteristic impacts speeds of ejecta are too low to efficiently produce additional ejecta which escape the satellites on impact. As a result, ejecta from one satellite colliding on another will not erode—it will accrete

Now recalling the ~$8 \times 10^{18}$(R/100 km)$^2$ gm total ejecta mass estimate from DS00, and pessimistically assuming lower limit radii of 20 km for these bodies (e.g., Weaver et al. 2006), we estimate that each of Nix and Hydra can eject sufficient material over the age of the solar system to coat one another equally to depths up to ~70 meters. However, since Nix and Hydra ejecta can also reach Charon and Pluto as well, one can calculate that if the debris reaches them in equal amounts that it reaches Nix and Hydra, then Pluto and Charon would respectively accumulate ~4 and ~14 cm of material from Nix and Hydra (combined) over time. These ejecta exchange depths are between ~$10^2$ and ~$2 \times 10^3$ times the required (~0.05 gm cm$^{-2}$) grammage of a layer sufficient to "paint" the native surface with its photometric properties, and will easily outpace the rate of surface impact overturning by impacts in 3.5 Gyr (<0.5 over-turnings in 4 Gyr; DS00). Hence, even if the transport process is two or three orders of magnitude inefficient, then Nix and Hydra will over time still cover one another and Pluto and Charon with ejecta derived from their regoliths. If the transport process is efficient, then the covering time for one another can be as short as a perhaps $3 \times 10^6$ yrs, perhaps about every $3 \times 10^8$ yrs for Charon, and 1 Gyr for Pluto.

Of course, a note of caution should be applied in the case of ejecta reaching Pluto Pluto. The narrow velocity window in which ejecta can reach Pluto but not escape the system implies a highly inefficient process and low accumulate amounts on Pluto. Moreover, Pluto's active surface-atmosphere deposition cycles (e.g., Stern et al. 1988; Spencer et al. 1997) will efficiently bury newly accreted material on the seasonal timescale, which is far shorter than the covering timescale. For both these reasons, as well as the narrow velocity window required for Nix/Hydra ejecta to reach Pluto but not escape the system, we do not expect Pluto's surface appearance to be much affected by material from Nix and Hydra.

No such ameliorating considerations affect Charon, Nix, and Hydra. As a result, we expect the surface colors, albedos, compositions, and phase properties of all three satellites to be first-order similar, as should be their surface temperatures. This is consistent with, and may explain why the colors of all three satellites are known to be similar (Stern et al. 2006). Albedo, phase curve, and temperature information are not available to test our predictions, but all of these parameters will be available after the New Horizons encounter with the Pluto system in 2015.

The propensity for Nix and Hydra to coat regoliths of Pluto's satellites raises two other points that merit discussion.

The first is that their lightcurve amplitudes may largely be the result of shape effects, since albedo variations across their surfaces should be minimal owing to the effects of this exogenous coating[2].

Second, the prediction of similar albedos on Charon, Nix, and Hydra also suggests that the sizes of Nix and Hydra can be predicted to be near the sizes corresponding to Charon-like albedos. Taking the measured p~0.35 of Charon (Tholen & Buie 1997), one therefore derives diameter estimates of 44 km and 53 km for Nix and Hydra, respectively. When first order accurate mass determinations of Nix and Hydra emerge from improved four body orbit solutions, it will also be possible to compare their estimated bulk densities based on the assumption of Charon-like albedos made here. These radii and density predictions can also be tested by New Horizons.

**Implications for Satellite-Bearing Kuiper Belt Objects and Asteroids**

Noll et al. (2008) report that many binaries appear to display similar parent-satellite colors, though the dataset is sparse and error bars remain significant owing to PSF blending in many cases. Here we briefly examine whether the ejecta exchange mechanism operating in the Pluto system can also operate in asteroid and KBO systems with satellites.

One expects ejecta exchange to be efficient in a given system if:

1. $V_{ej}$ does not exceed the speed required to escape the system entirely, and,
2. The characteristic ejecta velocity $V_{ej}$ exceeds the escape speed from at least one body in the system, and
3. $V_{ej}$ is sufficient to put ejecta from in the system onto crossing orbits within the system.

Like the Pluto system, typical impact speeds on KBOs are 1-2 km s$^{-1}$, producing typical $V_{ej}$'s of 0.01 km s$^{-1}$ to 0.2 km s$^{-1}$. To determine if the ejecta exchange mechanism operates efficiently in any given asteroid or KBO system, one must know the individual masses and orbital architecture of the bodies in the given system. At present, there are few KBO satellite systems which are very well determined and future work will require better orbital data to explore bound orbit transfer in such systems.

---

[2]Of course, recent impacts would produce local effects that may not yet have been covered by infalling debris, and which also provide "windows" into the native properties of the crusts of these individual bodies.

Instead we will examine the role of hyperbolic ejecta exchange in such systems. Although less efficient, estimates of its effects require less information about the bodies in the system and their orbits.

Hyperbolic ejecta exchange operates whenever $V_{ej}>V_{esc}$ where $V_{esc}$ is the velocity required to escape both the emitting satellite and to then escape the KBO from the orbit of the satellite. When this condition is satisfied, hyperbolic ejecta can strike a body in the system if its departure trajectory fortuitously intersects the position of that body.

We can approximate the magnitude of such exchange by recognizing that over time, the hyperbolic asymptotes of ejecta created by KBO impactors on KBOs and their satellites will be isotropic in the frame of the target system. Combining this fact with the $\sim 8 \times 10^{18}(R/100 \text{ km})^2$ gm total ejecta mass estimate discussed above, one can quickly show that for bodies that do not rotate synchronously relative to one another, an optically important 0.05 gm cm$^{-2}$ thick layer of impact ejecta will accumulate in the age of the solar system so long as the ratio of the separation between the two bodies in the primary to the radius of the eroding object, $D^*_{max} \sim 90$ target radii. In the case of synchronous rotation, this separation ratio constraint increases by 4 to $D^*_{max} \sim 360$. Hence, for example, for impact targets with a 100 km radius, the other body in the system must orbit closer than 9000 to 36000 km, depending on whether the two bodies are in synchronous rotation or not. Smaller separations will increase the expected depth of the exchange coating that will accumulate over time[3].

Using system parameters given in Noll et al. (2008), we can evaluate the effectiveness of this process for five KBO binaries where the system parameters are well determined by examining the total $V_{esc}$ from the parent (or satellite) bodies, and the $D^*$ values for both of these bodies as well, based on their radii. Table 3 provides the data to do so.

Table 3: Hyperbolic Ejecta Exchange Properties for Some KBO Binary Systems

| KBO Binary | $V_{esc}$ (Primary) | $D^*$ (Primary) | $V_{esc}$ (Satellite) | $D^*$ (Satellite) |
|---|---|---|---|---|
| Eris | 1351 m s$^{-1}$ | 30 | 247 m s$^{-1}$ | 230 |
| 2003 EL61 | 898 m s$^{-1}$ | 70 | 228 m s$^{-1}$ | 292 |
| **1999 TC36** | **102 m s$^{-1}$** | **40** | **39 m s$^{-1}$** | **111** |
| **2001 QC298** | **187 m s$^{-1}$** | **15** | **177 m s$^{-1}$** | **16** |
| **1998 SM165** | **83 m s$^{-1}$** | **85** | **26 m s$^{-1}$** | **280** |

Notes: The calculations that populated Table 3 were made assuming identical parent-satellite albedos and densities. Where $V_{esc}<200$ m s$^{-1}$ or $D^*<360$ we indicate these parameters in bold. Systems which satisfy both constraints should undergo significant hyperbolic mass exchange and have their names also indicated in bold.

---

[3]Of course, if the primary is sending material to a synchronous secondary, only the primary-facing hemisphere of the secondary will receive such hyperbolic ejecta. If a synchronous secondary is sending material to the primary, then said material will originate from the primary-facing hemisphere of the secondary; if in this case the system is in double synchronicity, then only the satellite-facing hemisphere of the primary will receive such hyperbolic ejecta

From these data, we find that hyperbolic ejecta exchange can occur in all of these KBO binaries, except for Eris and EL61 where $D^*$ is acceptable but $V_{esc}$ is too high (though even these bodies present borderline cases).

Based on these results, we predict that ejecta exchange can also cause the parent-satellite albedos, compositions, and phase curves to be similar in such systems[4].

Finally, let us consider the asteroid belt and NEO population. Here, impact speeds are typically several higher than in the KB. This is because these bodies orbit much closer to the Sun, so the characteristic impact speeds they experience are of order 3-5 km s$^{-1}$. However, Michel et al. (2004) report ejecta speeds not very different from those we estimate above in the KB. As a result, while bound ejecta may be rare, hyperbolic ejecta exchange may also result in like parent-satellite surface colors, albedos, etc. in some asteroid parent-satellite systems, as it does in the KB.

## Conclusions

We have shown that ejecta regolith exchange occurs efficiently in the Pluto system, and is likely the reason for the similar colors of Charon, Nix, and Hydra. Based on this, we have predicted that Charon, Nix, and Hydra also share like albedos, colors, compositions, phase curves, and surface temperatures due to ejecta exchange. We have also predicted that the expected sizes of Nix and Hydra can be best estimated by assuming Charon-like albedos.

We have further shown that the ejecta exchange is not specific to the Pluto system, but will also operate in other KB satellite systems, but that it is unlikely to be as effective in asteroid binary systems.

A more thorough investigation of this process involving actual Monte Carlo impact simulations on specific KB and asteroid binaries, followed by ejecta orbit integrations, is planned to further explore this process.

## Acknowledgements.

---

[4]The concern that much more mass will be lost from a typical KB or KB satellite owing to its own erosion by KB bombardment than the mass it receives from its partner body through hyperbolic ejecta exchange is not concerning in the DS00 show that for a typical 100 km radius KBO, only 7-32% of the body's surface will be covered in craters over the age of the solar system. So one still expects areal mixing to be dominated by the high fraction of surface that is not eroded by KB bombardment, but which will have received hyperbolic ejecta from the partner body.

I thank Luke Dones, Dan Durda, Will Grundy, Keith Noll, and Hal Weaver for valuable comments on this manuscript. This paper is Lunar and Planetary Institute contribution 413.